\documentclass[epj]{svjour}

\usepackage{graphicx,amssymb,amsmath}
\usepackage[applemac]{inputenc}

\begin{document}

\title{Fourier analysis of wave turbulence in a thin elastic plate}
\author{Nicolas Mordant}
\institute{Laboratoire de Physique Statistique, Ecole Normale Sup\'erieure \& CNRS, 24 rue Lhomond, 75005 Paris, France.}


\abstract{
The spatio-temporal dynamics of the deformation of a vibrated plate is measured by a high speed Fourier transform profilometry technique. The space-time Fourier spectrum is analyzed. It displays a behavior consistent with the premises of the Weak Turbulence theory. A isotropic continuous spectrum of waves is excited with a non linear dispersion relation slightly shifted from the linear dispersion relation. The spectral width of the dispersion relation is also measured. The non linearity of this system is weak as expected from the theory. Finite size effects are discussed. Despite a qualitative agreement with the theory, a quantitative mismatch is observed which origin may be due to the dissipation that ultimately absorbs the energy flux of the Kolmogorov-Zakharov casade.
}

\maketitle

The kinetic Weak Turbulence Theory (WTT) describes the long time statistics of a large number of waves exchanging energy through a weak non linear coupling. It has been applied to a vast number of systems such as Alfven waves in solar wind, oceanic surface or internal waves, optics in non linear media, Bose-Einstein condensates... (see \cite{Zakharov,Newell} for a collection of examples). It predicts in particular a Kolmogorov-Zakharov energy cascade as a stationary solution of the non equilibrium case. Energy is transferred conservatively from the large forcing scales down to small scales (at which it is dissipated) through multiple waves resonances. In some cases, an inverse cascade can be observed if the wave action is conserved as well. This phenomenology shares many similarity with hydrodynamical turbulence (either in 2D or 3D) so that information provided by the WTT could be of interest in the quest of the still missing statistical theory of fluid turbulence. 

The main hypotheses of the WTT are (i) a infinite size system and (ii) weak non linearity of the waves. (i) ensures that frequencies are not discrete so that resonance conditions can be fulfilled. (ii) is required in the theory so that to have scale separation between linear (fast) time scales and non linear (slow) time scales. In laboratory experiments, (i) can obviously not be achieved rigorously and waves have finite amplitudes so that the validity of both (i) and (ii) can be discussed. This is especially the case for gravity surface waves in which both hypotheses could be non valid simultaneously~\cite{Denissenko,Erics}. Another concern in experiments is dissipation. The theory requires an additional hypothesis: (iii) the energy sinks are supposed to act at scales widely separated from the forcing so that energy flows conservatively in a inertial range of scales. This condition is often difficult to achieve in experiments, let alone in numerics. This may also alter the statistics of the waves~\cite{Connaughton}. Additional hypotheses are (iv) locality of interactions in $k$-space and (v) phases of interacting modes are random.

In many cases, measurements are difficult to implement as in measurements of solar winds or in confined plasmas. So far measurements have been restricted mostly to spectral exponents. The measurements are often performed at a single point in space so that only temporal Fourier spectra can be accessed. Here we use a laboratory experiment which setup is sufficiently simple so that both the temporal and the spatial evolution of wave turbulence can be measured: wave turbulence in a thin elastic plate. This system has interests in its own (cymbals, vibration of structures in aeronautics) but also as a toy model of wave turbulence. The WTT has recently been applied to this system~\cite{During} and predicts 4-wave resonances. Experiments show a disagreement between the measured single point spectrum and the theoretical prediction~\cite{Boudaoud,Mordant}. The application of a high speed profilometry technique provides a measurement of the dynamics of the waves resolved both in time and space over a significant portion of the plate~\cite{Cobelli}. It is possible for the first time to get a measurement of the spatio-temporal structure of wave turbulence and thus to test the WTT on more elaborate predictions as just single point spectral exponents.

In the following, we recall previous theoretical and experimental results, then we present shortly the experimental setup. We analyze the statistics of the wave amplitude and the various Fourier spectra of the deformation. Finally we discuss the issue of finite size effects.

\section{Summary of previous results}

\subsection{Theoretical results}

The evolution equations for the deformation of a thin elastic plate are the F\"oppl-Von K\'arm\'an equations:
\begin{eqnarray}
\rho\partial_{tt}\zeta+\frac{h^2E}{12(1-\sigma^2)}\Delta^2\zeta=\{\zeta,\chi\}\label{le}\\
\frac{1}{E}\Delta^2\chi=-\frac{1}{2}\{\zeta,\zeta\}\label{nle}
\end{eqnarray}
where $\zeta$ is the normal displacement and $E\simeq2\times10^{11}Pa$ is the Young modulus, $\sigma\simeq0.3$ is the Poisson ratio, and $\rho\simeq8000$~kg.m$^{-3}$ is the specific mass (values are typical for stainless steel). $\Delta$ is the 2D Laplacian, and $\{a,b\}=\partial_{xx}a\,\,\partial_{yy}b+\partial_{yy}a\,\,\partial_{xx}b-2\partial_{xy}a\,\,\partial_{xy}b$. $\chi$ is the stress function and is related to the stretching of the plate~\cite{Landau}.

These equations support linear dispersive waves (lhs terms in (\ref{le})) whose linear dispersion relation (LDR) is 
\begin{equation}
\omega_L(k)=\sqrt{\frac{Eh^2}{12(1-\sigma^2)\rho}}k^2\, .
\label{ldr}
\end{equation}
 For large amplitudes, the wave equation exhibits cubic nonlinearities (rhs of (\ref{le}) and (\ref{nle})). These nonlinear equations are believed to lead to wave turbulence \cite{During}. The WTT has been applied to the case of thin elastic plates by D\"uring {\it et al.}~\cite{During}. Starting from (\ref{le}) and (\ref{nle}), the waves are supposed to be weakly non linear. This enables a slow enough transfer of energy among waves so that the wave amplitudes evolve on much longer time scales than the linear period~(\ref{ldr}). In that case a stationary out of equilibrium regime can be exhibited in which energy is transferred through 4-wave resonances from large scales to small scales. In the following, we refer to ``space spectrum" for a power spectrum density $E(\mathbf k)$ using the $\mathbf k$ (or $k=\|\mathbf k\|$) variable, to ``time spectrum" for a power spectrum density $E(\omega)$ using the $\omega$ variable (or $f=\omega/2\pi$) and ``space time spectrum" for $E(\mathbf k,\omega)$ (or $E(k,\omega)$). We add a $\zeta$ or $v$ ($v=\partial_t \zeta$) subscript to $E(\dots)$ to specify which quantity is used (deformation or normal velocity). The WTT predicts the following space spectrum for the displacement $\zeta$:
\begin{equation}
E_\zeta(k)=C\frac{P^{1/3}}{(1-\sigma^2)^{1/6}}\frac{\ln^{1/3}(k^\star/k)}{\sqrt{E/\rho}k^3}\, ,
\label{during}
\end{equation}
where $P$ is the energy flux per unit mass (here we use the definition of Connaughton {\it et al.}~\cite{Conn}), $C$ a dimensionless number, and $k^\star$ is a cutoff wave number. Using (\ref{ldr}) to change variables from $k$ to $\omega$, the theoretical prediction for the velocity time spectrum is then
\begin{equation}
E_v(\omega)=C^{\prime}\frac{hP^{1/3}}{(1-\sigma^2)^{2/3}}\ln^{1/3}(\omega^\star/\omega)\, ,
\label{tsv}
\end{equation}
where $\omega^\star=\omega_L(k^\star)$ and $C^{\prime}$ is a number.

\subsection{Experimental results}
The motion of thin steel plates has been studied by Boudaoud {\it et al.}~\cite{Boudaoud} and Mordant~\cite{Mordant} using single or differential measurement of the velocity with a laser vibrometer. Single point measurements provide only the time spectrum. The experimental velocity spectra have be shown not to obey the analytical prediction (\ref{tsv}) neither in the frequency scaling nor in the energy flux scaling. The observed scaling is $E_v(\omega)\sim P^{0.7}\omega^{-0.6}$ rather than $E_v(\omega)\sim P^{1/3}\omega^{0}$. A continuous spectrum was observed at the highest forcing and differential measurements are consistent with a superposition of propagating sine waves whose dispersion relation is close to the LDR (\ref{ldr})~\cite{Mordant}. These observations rule out the possibility of strongly non linear structures being dominant in the motion of the plate as suggested by Boudaoud~{\it et al.}~\cite{Boudaoud}. 

The reason for this disagreement between theory and experiment remains an open question. Among the possibilities, one can list:
\begin{enumerate}
\item finite size effects: these induce a quantization of the frequencies of the normal modes of the plate. If the spectral (frequency) width of the modes is not wide enough this could lead to discrete wave turbulence or laminated wave turbulence. The discretization of the frequencies may hinder the occurrence of 4-wave resonances and alter the energy flux~\cite{Kartashova,Kartashova1,Kartashova2}.
\item deformation of the plate at rest: if the plate is not infinitely flat at rest (which is always the case to some extend for real plates) then the non linear terms of the evolution equations may change from cubic to quadratic and this would also alter the energy transfer through 3-wave resonances in place of 4-wave~\cite{Boudaoud}.
\item dissipation: if the dissipation is not localized at high frequency as assumed in the WTT, then it would alter the energy transfer. Energy would only be partially transferred among scales. Note that the source of dissipation here is not obvious and can involve the bulk material dissipation, radiation of acoustic waves, transfer of energy to the ambient air or to the support... It has been discussed to some extent in \cite{Boudaoud}.
\end{enumerate}

More recently, a Fourier Transform Profilometry (FTP) technique has been applied to this problem by Cobelli {et al.}~\cite{Cobelli}. This technique, coupled with a high speed camera, enables the measurement of the deformation of a significant portion of the surface of the plate, resolved in time and space. It was then possible to compute the space-time $(\mathbf k,\omega)$ spectrum of the deformation. Here we report more extensive measurement and analysis using the FTP technique with an upgraded camera.

\section{Description of the experiment and the data processing}

\subsection{Experimental setup}

\begin{figure}[!htb]
\centering
\includegraphics[width=7cm]{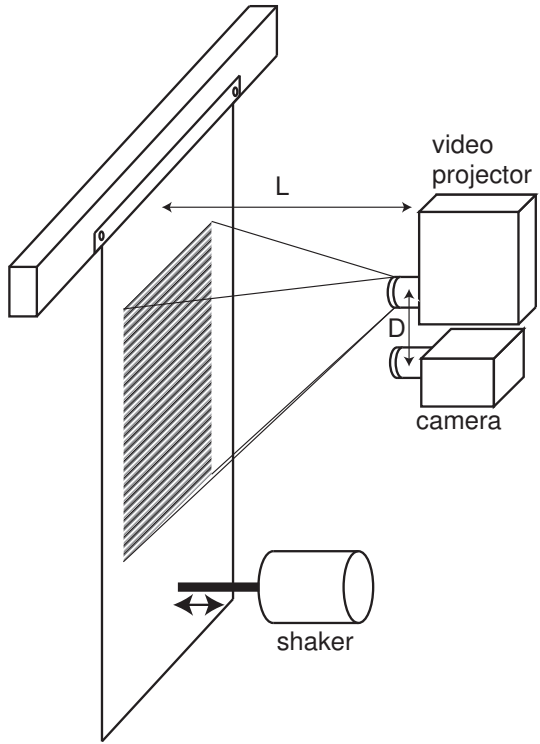}
\caption{Schematics of the experiment: the plate is $2\times1$~m$^2$ and hangs under its own weight under a beam. The motion is forced by a electromagnetic shaker excited by a sine current at 30~Hz which amplitude can be tuned. A high definition video projector projects a sine intensity pattern on the plate. The deformation of the plates induces a deformation of the pattern seen by a high speed camera. The field of view of the camera is about $0.8\times1.1$~m$^2$.}
\label{setup}
\end{figure}
The experimental setup is similar to that described in \cite{Mordant,Cobelli}. The plate is a stainless steel plate of size $2\times1$~m$^2$ and thickness 0.4~mm. It is bolted on a beam at the top along one short edge. It hangs freely under its own weight. At 40~cm above the bottom, a electromagnetic shaker is anchored on the plate and forces its motion (see schematics in fig.~\ref{setup}). The excitation current is a pure sine at 30~Hz. Its amplitude can be tuned to explore various forcing intensities. The average power transmitted to the plate by the vibrator has been observed to be quadratic in the amplitude of the excitation current~\cite{Boudaoud}. Here we measure the average input power through arbitrary units of the forcing current in the interval [0.5,6] corresponding to a average power spanning an interval proportional to [0.25,36], i.e. two orders of magnitude.

The FTP technique requires the projection of a sine intensity pattern $I(x,y)\propto \sin(2\pi x/p)$ on the plate~\cite{Cobelli1,Maurel}. The plate is painted matte white to behave as a diffusive screen. Here $x$ is the vertical coordinate. The wavelength of the pattern on the plate is $p=6.6$~mm. The pattern is projected by a High Definition Epson video projector. The pattern is recorded by a Photron SA1 high speed camera at a frame rate reaching 8000 frames per second (fps) at the highest forcing. The field of view of the camera is about $0.8\times1.1$~m$^2$. This region is located about 10~cm away from the edges and in the central region between the top of the plate and the anchor point of the shaker in order to avoid possible boundary inhomogeneities. The field of view is significantly larger than reported in~\cite{Cobelli} and the camera is both faster and more sensitive as that used previously.
The FTP scheme is identical to that of ~\cite{Cobelli,Cobelli1}. The projector and camera lenses have parallel optical axes. They are $L=2.17$~m away from the plate at rest and the separation of the optical axes is $D=30$~cm. 

\subsection{Data processing}

\begin{figure}[!htb]
\centering
(a)\includegraphics[width=9cm]{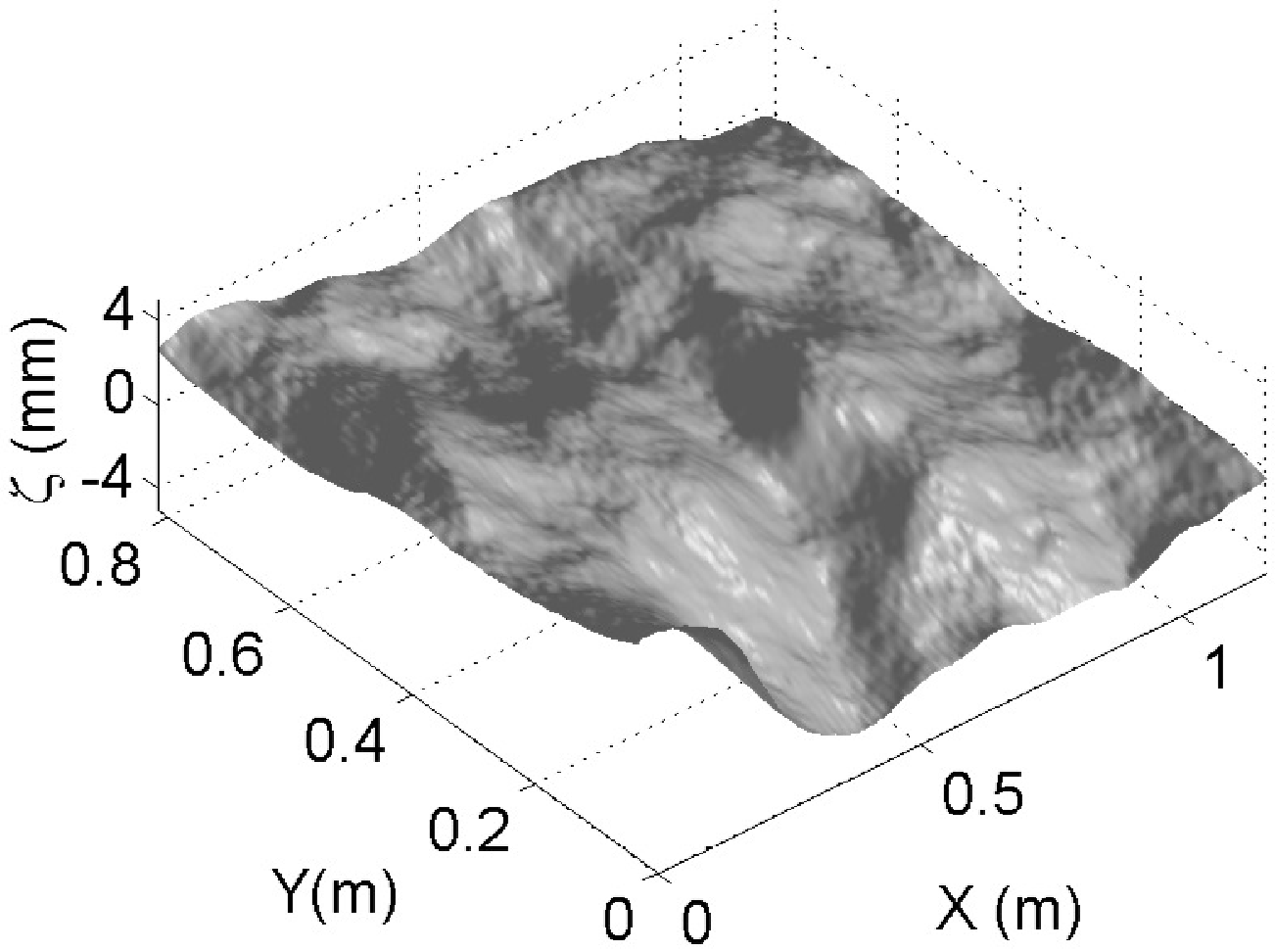}
(b)\includegraphics[width=9cm]{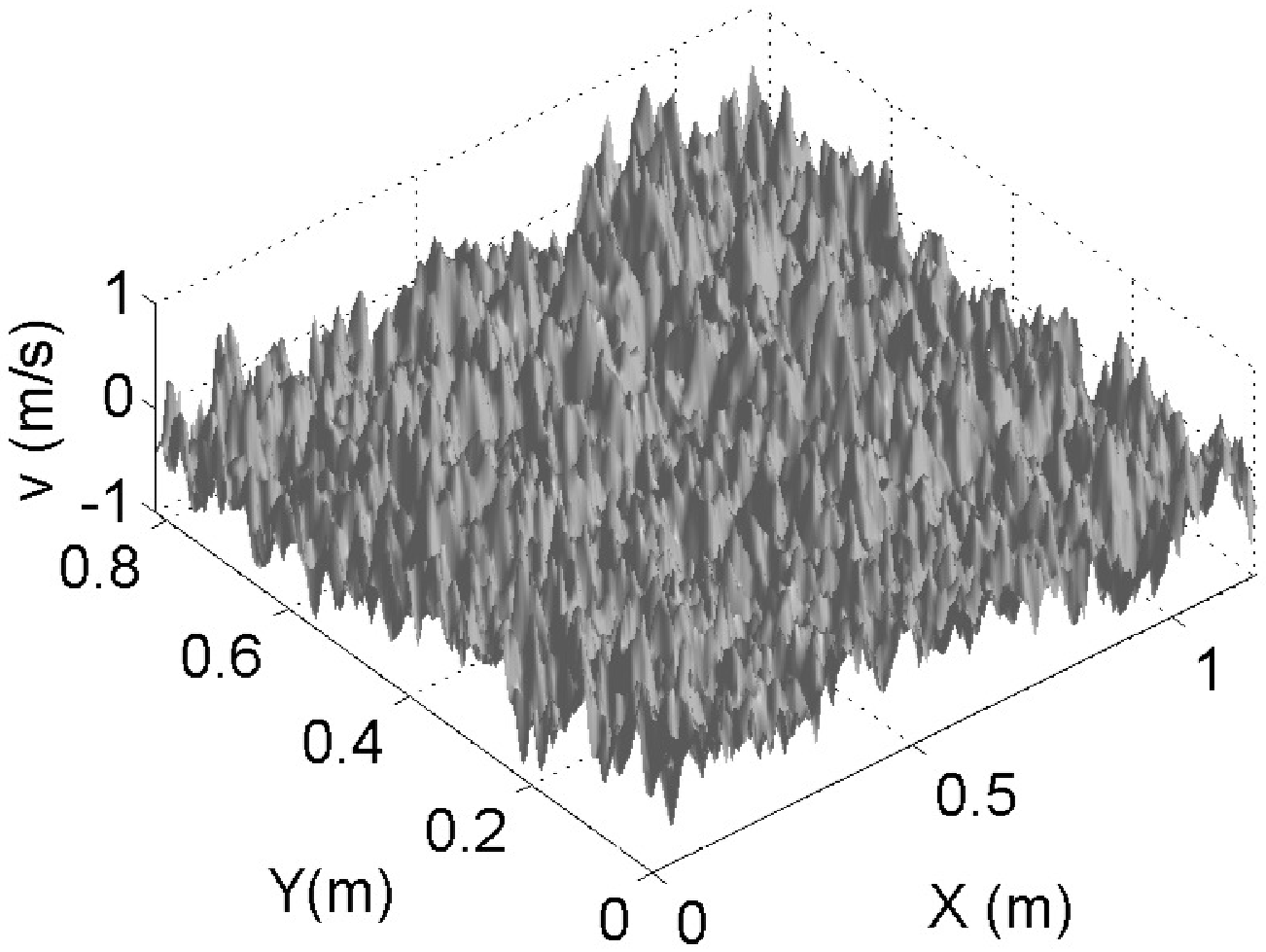}
\caption{Example of the plate deformation and velocity}
\label{example}
\end{figure}

The image processing is similar to that described in~\cite{Cobelli1}. The deformation of the plate appears as a phase modulation of the pattern recorded by the camera. The phase of the deformed pattern is extracted and the deformation is computed as in~\cite{Cobelli1}. In this way, one obtains a movie of the deformation of the plate over about one square meter. At each forcing amplitude, the recording rate of the camera is tuned to ensure that the Shannon criterium is satisfied up to frequencies at which the signal falls below the noise level. The velocity is computed by difference between successive deformation frames. A snapshot of deformation and velocity is shown in fig.~\ref{example}. The spectrum of the deformation being quite steep (about $1/k^4$) the deformed surface is quite smooth. On the contrary the spectrum of the velocity is quite flat (see below) resulting in a velocity that varies very quickly in space.

From the deformation or velocity movies many analyses can be performed. As shown below, a Fourier transform in space can be applied to get the spectral wave amplitude $v(\mathbf k,t)$. This component is the sum of $v^+(\mathbf k,t)$ and $v^-(\mathbf k,t)$ respectively the wave components with positive frequency $\omega>0$ and the wave component with negative frequency (these two waves are traveling in opposite directions). These two components can be separated numerically. By symmetry one gets $v^+(\mathbf k,t)=v^-(-\mathbf k,t)^\star$ (where $^\star$ stands for the complex conjugate) so that one only has to study $v^+(\mathbf k,t)$ over the full complex plane to get all the information on waves traveling in all directions.
Statistics of these complex amplitudes (distributions, correlations...) can be computed. For instance, the 2D space spectrum $E_v(\mathbf k)$ is simply the average $\langle |v^+(\mathbf k,t)|^2\rangle$ where the average is taken over time. By integrating the 2D spectrum over angles of $\mathbf k$ one gets the 1D space spectrum $E_v(k)$. A Fourier transform in time can be applied also, leading to the spectral wave amplitude $v^+(\mathbf k,\omega)$. The space-time Fourier spectrum $E_v(\mathbf k,\omega)$ is obtained from the average $\langle |v^+(\mathbf k,\omega)|^2\rangle$ and it can also be integrated over the directions of $\mathbf k$ to get $E_v(k,\omega)$.

\section{Spectral wave amplitudes}

\subsection{Distribution of the wave amplitudes}

\begin{figure}[!htb]
\includegraphics[width=9cm]{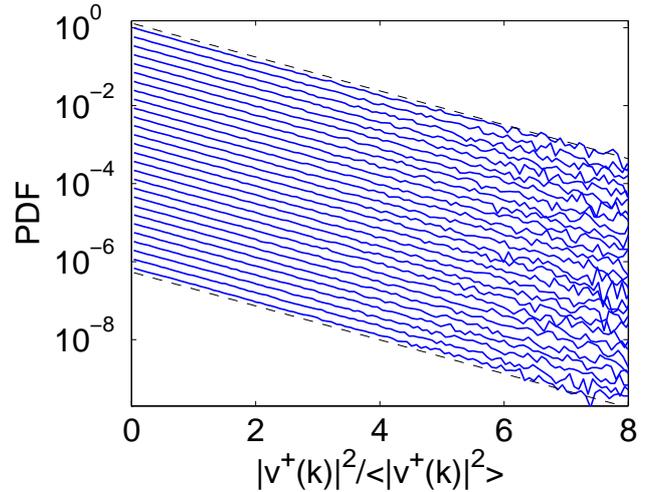}
\caption{PDF of the squared magnitude of the spectral deformation components $v^+(\mathbf k,t)$. Here, $\mathbf k=(k_x,0)$ with values of $kx$ equally spaced between 2.2 and 25. The various curves have been shifted vertically for clarity. Dashed lines correspond to exponential decay.}
\label{PDFamp}
\end{figure}
\begin{figure}[!htb]
\centering
\includegraphics[width=7cm]{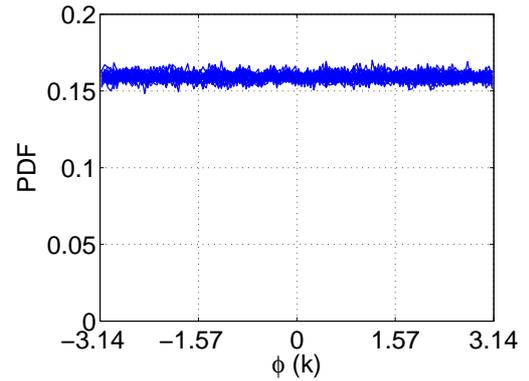}
\caption{PDF of the phase of the spectral deformation components $v^+(\mathbf k,t)$. Here, $\mathbf k=(k_x,0)$ with values of $kx$ equally spaced between 2.2 and 25. All curves are superimposed.}
\label{PDFphi}
\end{figure}

Figure~\ref{PDFamp} displays the measured distribution of the squared magnitude $|v^+(\mathbf k,t)|^2$ for values of $k$ spanning the interval between the forcing wavenumber up to wavenumbers corresponding to a signal over noise ratio close to 1. The distribution is exponential (Rayleigh distribution) as expected for a gaussian complex amplitude $v(\mathbf k,t)$. No departure is visible within the range of probabilities accessible with the current dataset, neither at small $k$ nor at large values. Figure~\ref{PDFphi} shows the distribution of the phases of $v(\mathbf k,t)$ for the same set of values of $\mathbf k$. The phases are corrected from the linear frequency, i.e. a linear drift $\omega_L(\mathbf k) t$ is subtracted. The distributions are flat at all values of $k$.

\begin{figure}[!htb]
\includegraphics[width=9cm]{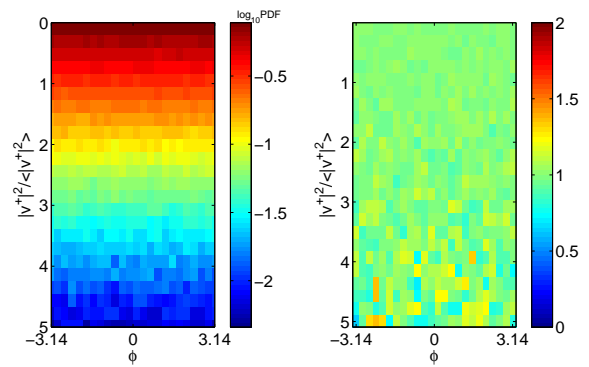}
\caption{example of the joint PDF of the magnitude and the phase of $v^+(\mathbf k,t)$ for $\mathbf k/2\pi=(15,0)$. Left: $P(\phi,|v^+|^2)$, right: ratio $P(\phi,|v^+|^2)/P(\phi)P(|v^+|^2)$.}
\label{jPDF}
\end{figure}
The phase and the magnitude of $v^+(\mathbf k,t)$ are also expected to be independent variables. Figure~\ref{jPDF} analyses the joint distribution of the magnitude and the phase for a given value of $\mathbf k$. The left part of the figure shows the joint distribution $P(\phi(\mathbf k,t),|v^+(\mathbf k,t)|^2)$ (where $\phi(\mathbf k,t)$ stands for the phase of $v^+(\mathbf k,t)$) and the right part shows the ratio $$\frac{P(\phi(\mathbf k,t),|v^+(\mathbf k,t)|^2)}{P(\phi(\mathbf k,t))P(|v^+(\mathbf k,t)|^2)}.$$
This ratio should be unity for independent phase and magnitudes. This is indeed observed in our experiment.
Note that to test the hypothesis (v) on random phases of interacting modes, one would have to use higher order statistical tools such as bi- and tricoherence.

\subsection{Correlation of the wave amplitudes}
\begin{figure}[!htb]
\includegraphics[width=9cm]{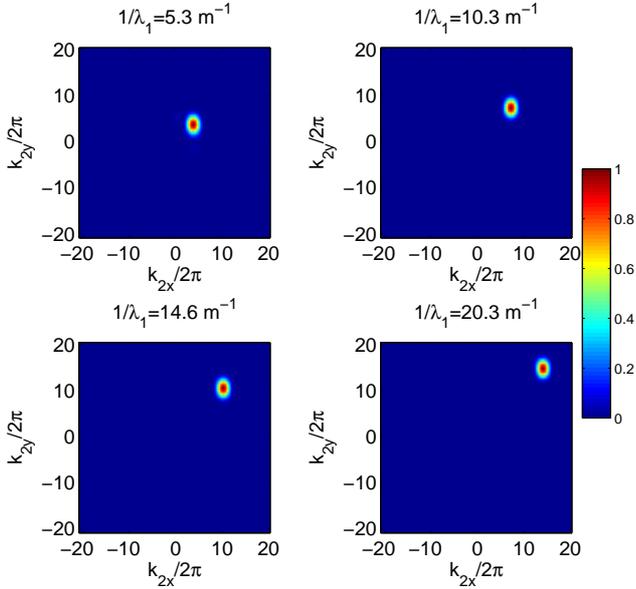}
\caption{Magnitude of the correlations $\left|\frac{\langle v^+(\mathbf k_1,t)v^+(\mathbf k_2,t)^\star\rangle}{\sqrt{\langle v^+(\mathbf k_1,t)^2\rangle \langle v^+(\mathbf k_2,t)^2\rangle}}\right|$. Here, $\mathbf k_1$ is chosen on the diagonal with its norm given in the title. }
\label{cor}
\end{figure}
Another test of the validity of the bases of the WTT is the fact that the spectral component should be uncorrelated i.e. 
\begin{equation}
\langle v^+(\mathbf k_1,t)v^+(\mathbf k_2,t)^\star\rangle\propto E_v(\mathbf k_1)\delta(\mathbf k_1-\mathbf k_2)
\end{equation}
This relation is tested in figure~\ref{cor}. The correlation are indeed zero except on $\mathbf k_1$. The width of the correlation peak is due to the finite resolution of the space Fourier transform due to the size of the recorded picture.

\section{The space-time Fourier spectrum}

\subsection{The Fourier spectra}
\begin{figure}[!htb]
\includegraphics[width=9cm]{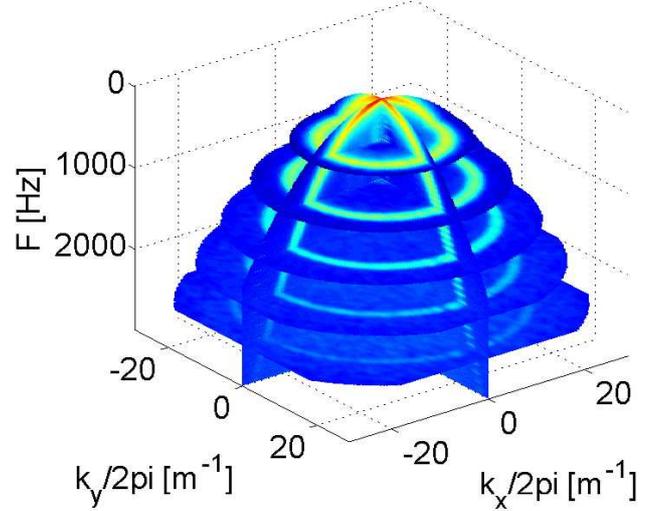}
\caption{The space-time spectrum $E(\mathbf k,\omega)$ for $P=16$. 3D view of the spectra with cuts at $k_x=0$, $k_y=0$ and $f=500$, 1000, 1500, 2000 and 2500~Hz. Energy levels are color coded in log scale.}
\label{sp3D}
\end{figure}
\begin{figure}[!htb]
\centering
\includegraphics[width=9.5cm]{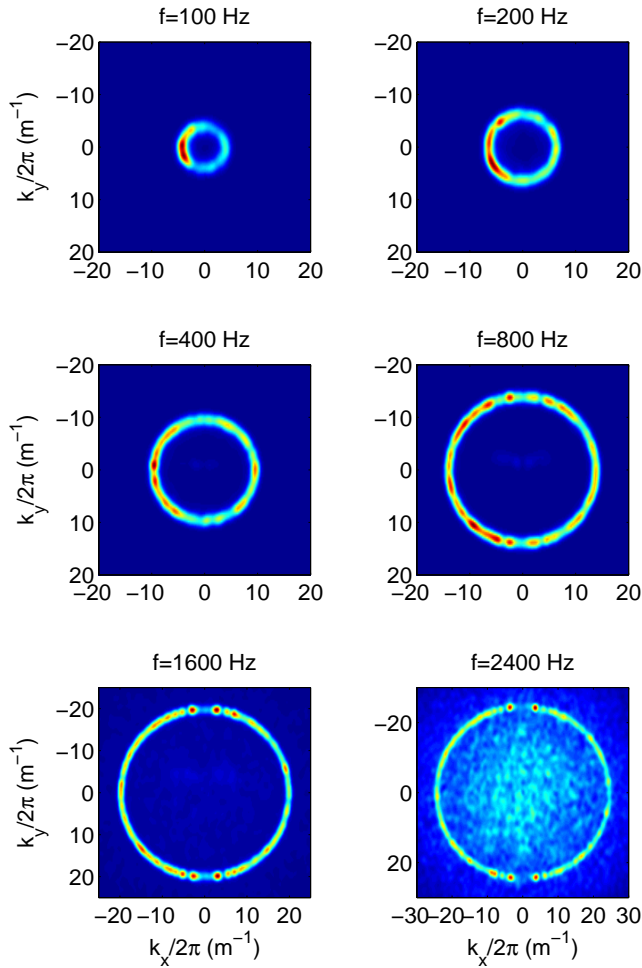}
\caption{The space-time spectrum $E(\mathbf k,\omega)$ for $P=16$: cuts of the spectrum at given values of $f$ (specified in the title of each subfigure).}
\label{spcut}
\end{figure}
As the full space and time resolved deformation can be measured using the FTP technique, we gain access to various Fourier spectra. Figure~\ref{sp3D} shows the full space and time spectrum $E_v(\mathbf k,\omega)$. It is very clear in this figure that energy is localized on a surface in the $(\mathbf k,\omega)$ space. This is what is expected for waves with a dispersion relation. The cuts of $E_v(\mathbf k,\omega)$ at constant $\omega$ (fig.~\ref{spcut}) shows that energy is localized on circles. The spectrum is shows some anisotropy at small $\omega$ due to the forcing that propagates waves preferentially towards the top of the plate. At higher frequencies, the energy is distributed continuously on the circles (except at the highest frequencies, see part~\ref{FSE} below) which shows that wave turbulence transfers energy in a continuous and almost isotropic manner in $\mathbf k$ space. 

\begin{figure}[!htb]
\includegraphics[width=9cm]{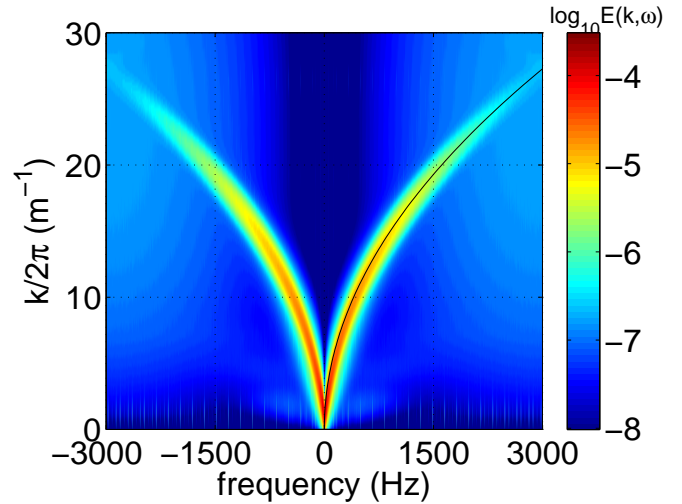}
\caption{The angle integrated space-time spectrum $E(k,\omega)$ for $P=16$.}
\label{sp2D}
\end{figure}
Figure~\ref{sp2D} displays the angle integrated spectrum $E(k,\omega)$. The linear dispersion relation is shown as the black line. The line of maximum energy is localized very close to the linear dispersion relation (this point is analyzed further in the next section).

\begin{figure}[!htb]
(a)\includegraphics[width=9cm]{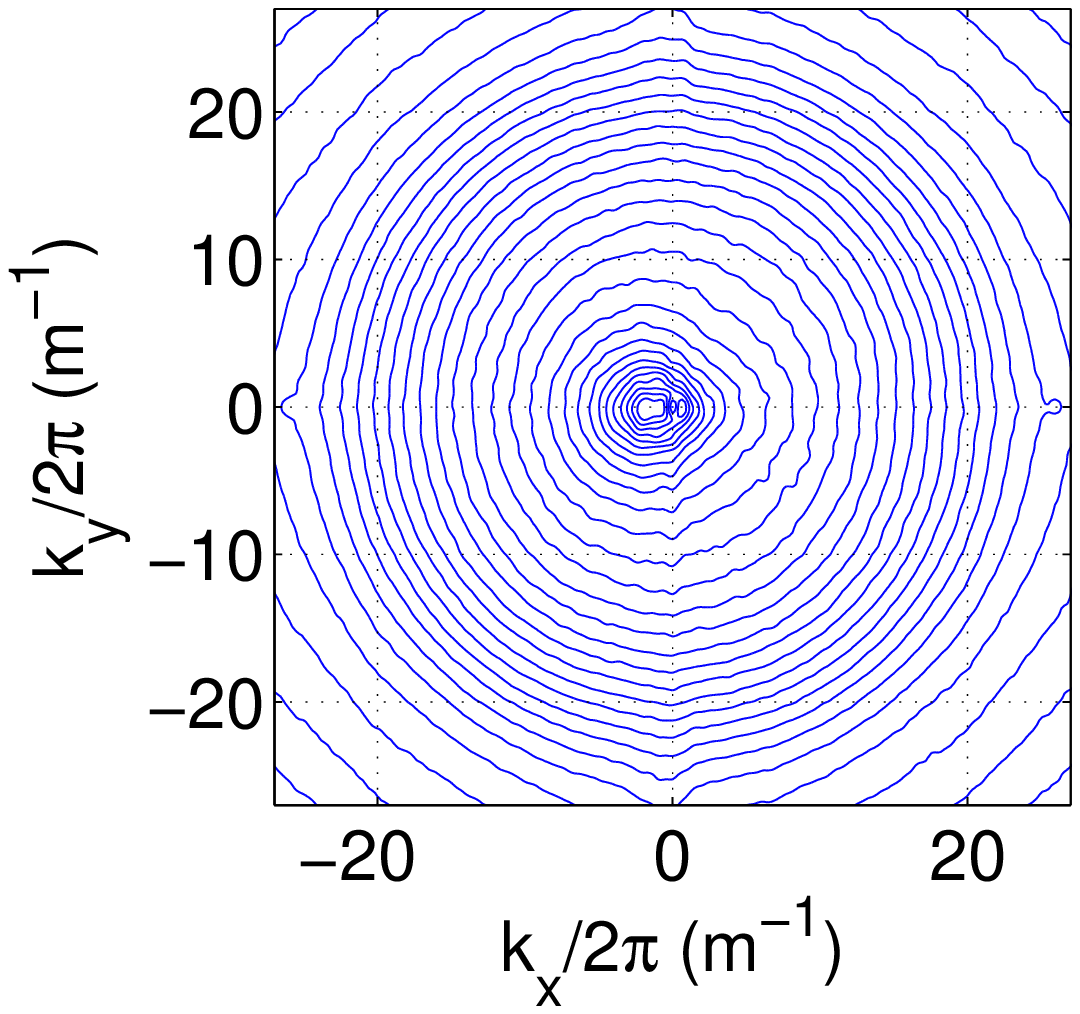}
(b)\includegraphics[width=9cm]{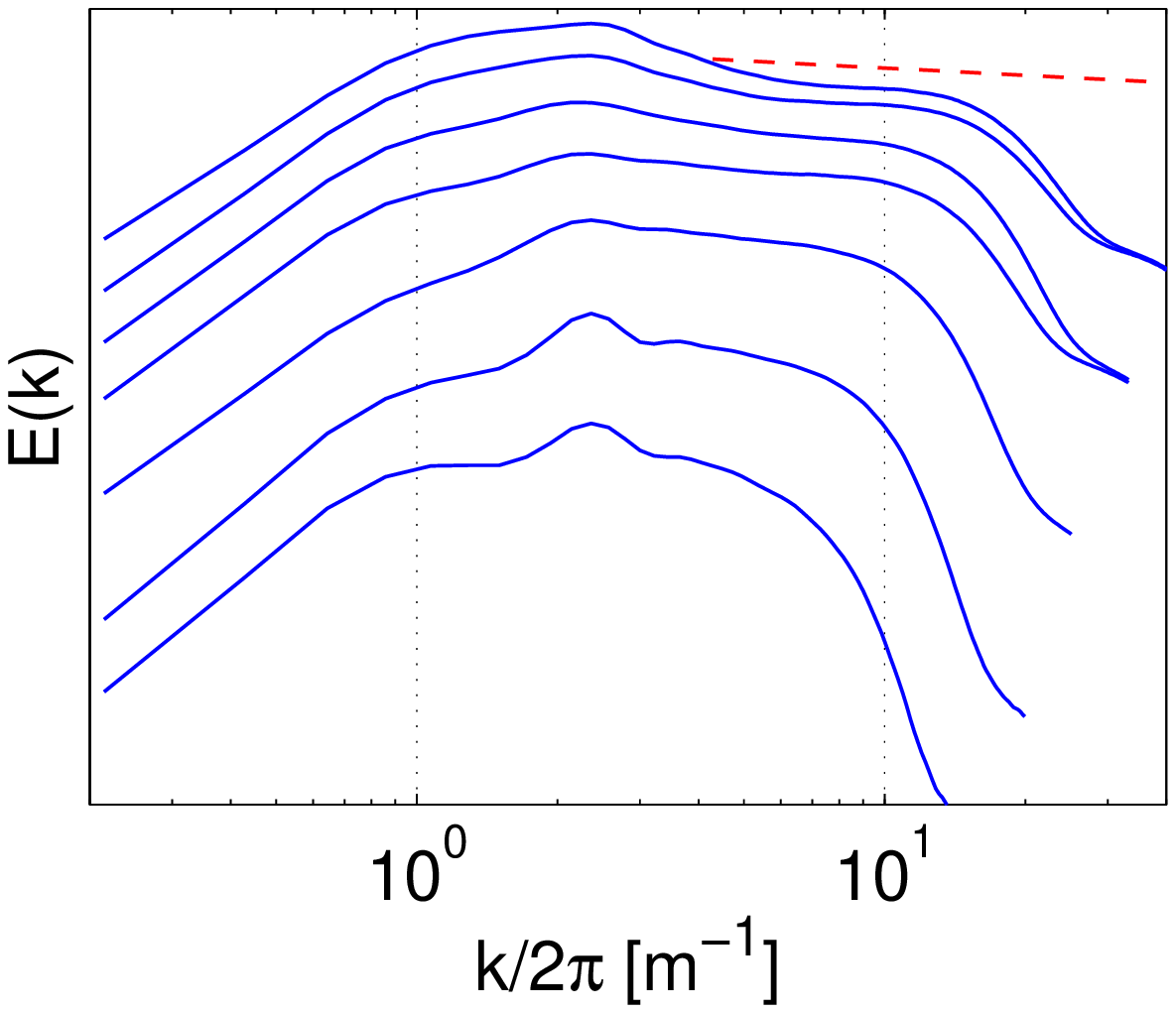}
\caption{The space spectra $E(\mathbf k)$ (a) for $P=16$ and the angle integrated space spectrum $E(k)$ (b) for $P=0.25$, 1, 4, 9, 16, 25 and 36 (from bottom to top). The dashed line is $k^{-0.1}$. The forcing is at $k=2.8$~$m^{-1}$.}
\label{speck}
\end{figure}
The isotropy of the space spectrum is clearly seen in figure~\ref{speck}(a). At low $k$ the anisotropy of the forcing is obvious but at higher $k$ the lines of constant energy are circles. Angle integrated spectra $E(k)$ are shown in fig.~\ref{speck}(b) for $P$ ranging from 0.25 to 36. The development of a short plateau (almost horizontal) can be seen that corresponds to an inertial regime. This inertial regime is not so wide so that hypothesis (iii) on scale separation of forcing and sinks can be questioned here.

\begin{figure}[!htb]
\includegraphics[width=9cm]{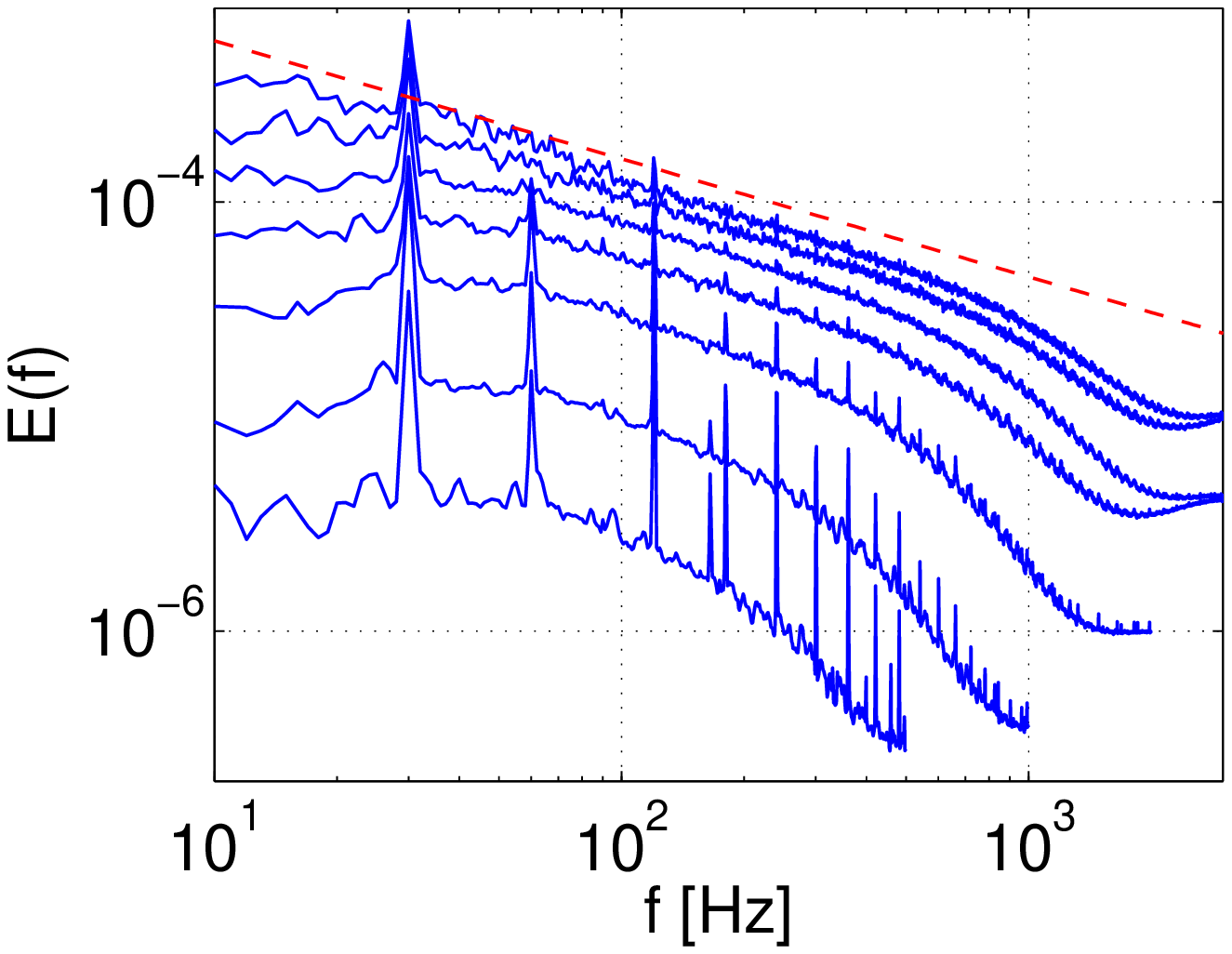}
\caption{The time spectrum $E(\omega)$ for $P=0.25$, 1, 4, 9, 16, 25 and 36 (from bottom to top). The spurious peaks in the spectra are either harmonics of the forcing or due to flickering of the videoprojector acting at small $k$ as also seen in fig.~\ref{sp2D}.}
\label{specf}
\end{figure}
This inertial regime is much wider in time spectra (fig.~\ref{specf}), due to the curvature of the dispersion relation $\omega\propto k^2$. A $f^{-0.55}$ region is clearly visible as was already reported in previous work~\cite{Boudaoud,Mordant}. Using the linear dispersion relation this translates into $k^{-0.1}$ as seen in fig.~\ref{speck}(b).

\subsection{The non-linear dispersion relation}

\begin{figure}[!htb]
\includegraphics[width=9cm]{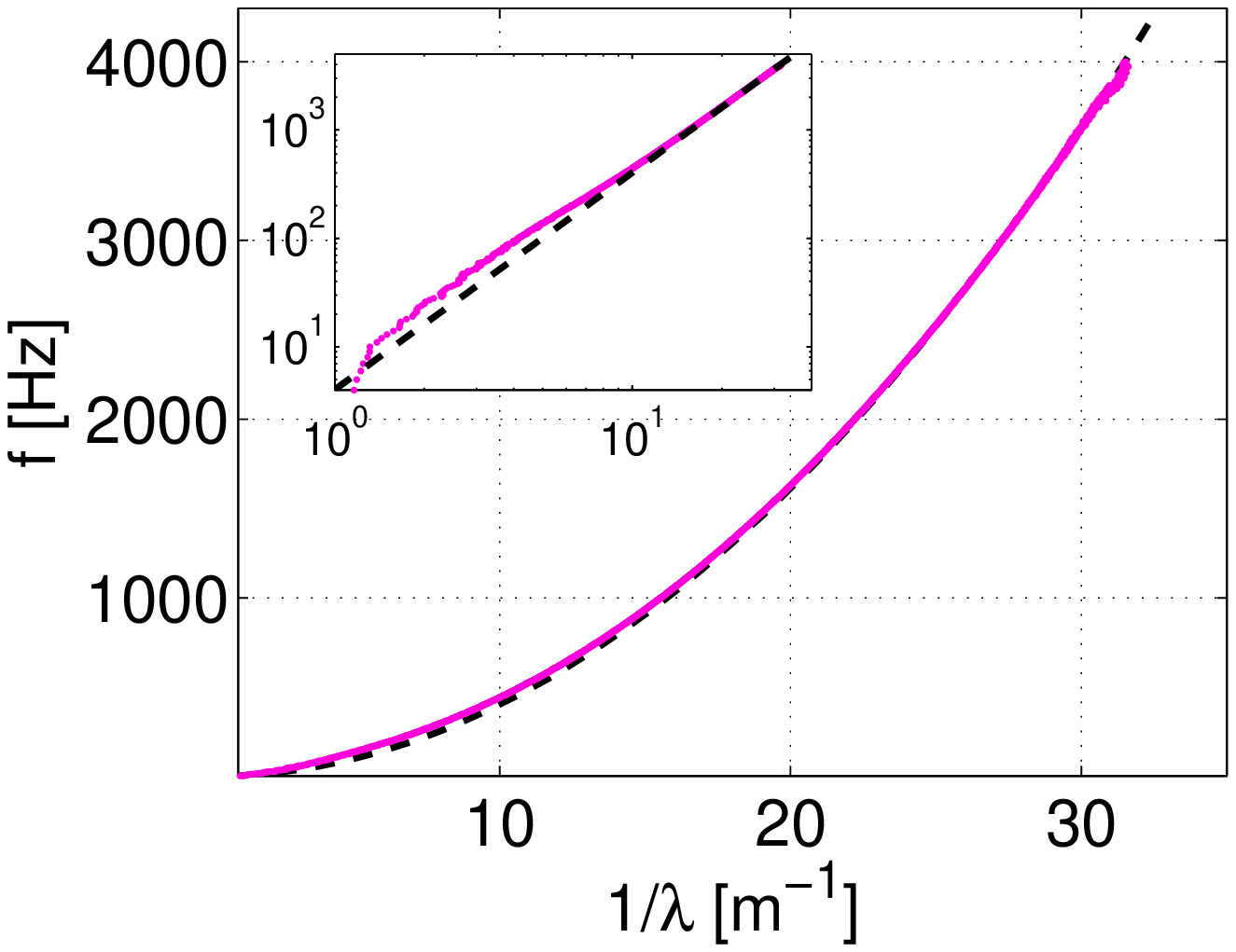}
\caption{The non linear dispersion relation at $P=36$. The insert is in loglog scale.}
\label{rd}
\end{figure}
A non linear dispersion relation (NLDR) can be obtained by extracting the position of the crest of the energy line in fig.~\ref{sp2D}. An example is shown in fig.~\ref{rd} at the highest forcing $P=36$. At high frequency it recovers the $k^2$ scaling of the linear dispersion relation in the range of scales in which the spectra are decaying at the fastest rate. This allows actually to compute very precisely the constant $C$ in the LDR $\omega_L=Ck^2$. Some uncertainty exists in the physical parameters $E$, $\sigma$ and so on which provides estimations of $C$ only within a few percent at best. Here we estimate $C=0.641$~m$^2$s$^{-1}$ which gives the best agreement with the high $k$ evolution of the NLDR. The accuracy on the estimation of $C$ is about $0.2\%$. 

\begin{figure}[!htb]
(a)\includegraphics[width=8.5cm]{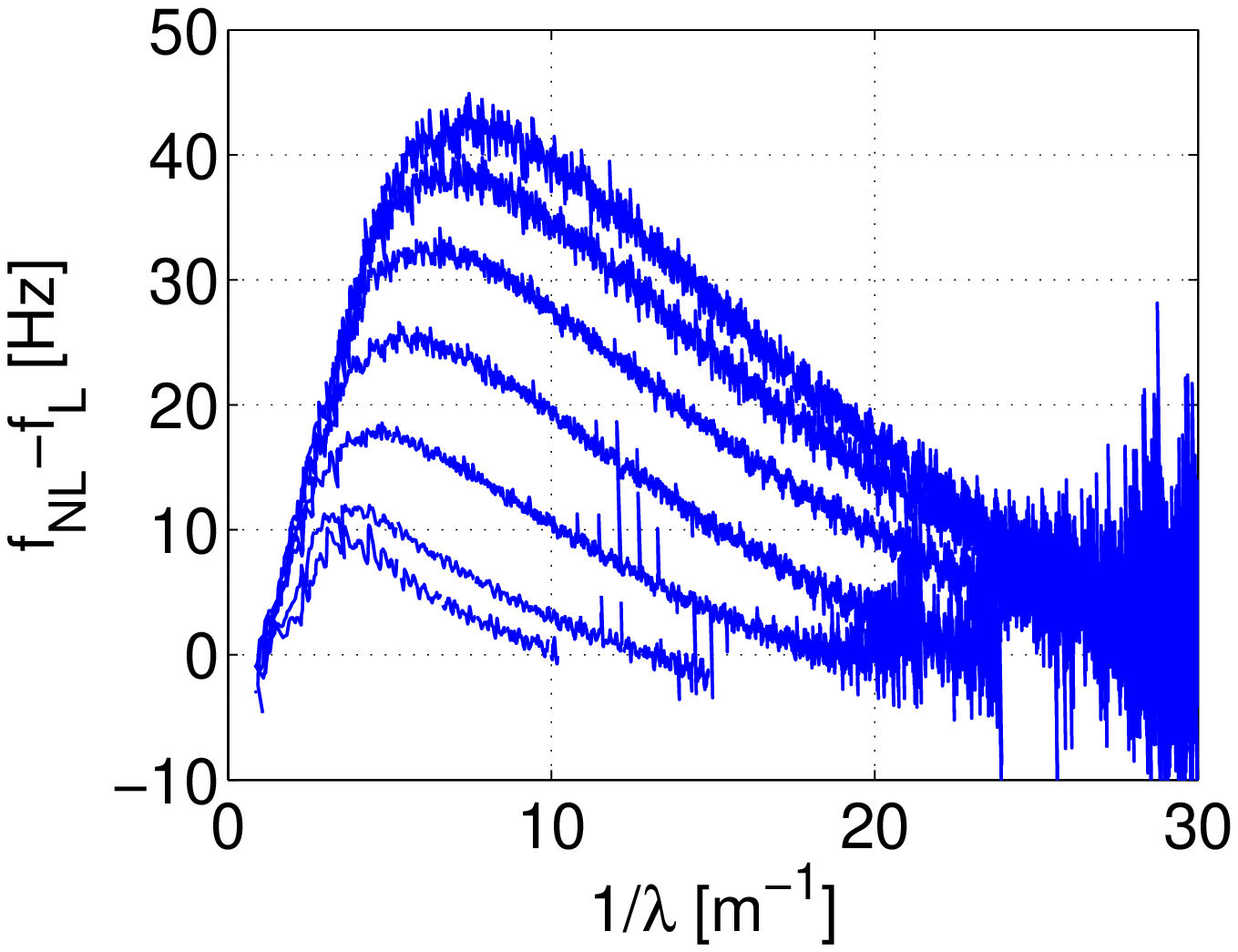}
(b)\includegraphics[width=8.5cm]{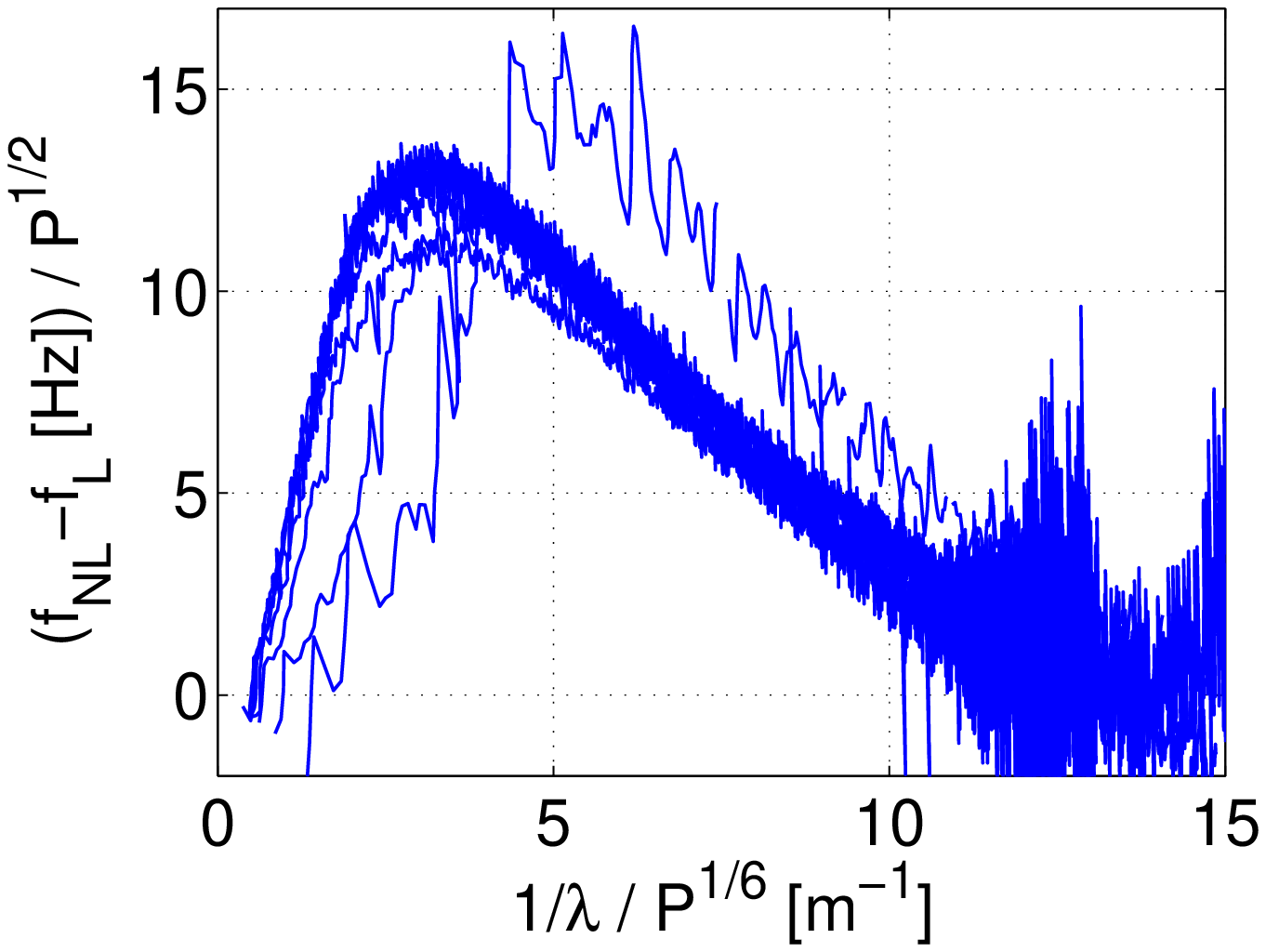}
\caption{(a) The shift of non linear dispersion relation in respect to the linear dispersion relation $\Delta \omega(k)=\omega_{NL}(k)-\omega_L(k)$ for $P=0.25$, 1, 4, 9, 16, 25 and 36 (from bottom to top). The forcing is at $k=2.8$~$m^{-1}$. (b) $\Delta \omega(k)/P^{1/2}$ versus $k/P^{1/6}$.}
\label{shift}
\end{figure}
The WTT predicts a small correction to the LDR under the form $\omega_{NL}(k)=\omega_L(k)+\Delta \omega(k)$~\cite{Newell}. The shift $\Delta \omega(k)$ is shown in fig.~\ref{shift}(a). It rises from zero and peaks at relatively low $k$ (but still larger than the forcing at $k=2.8$~m$^{-1}$. It decays to zero at the highest $k$ available (values of $k$ at which the signal over noise ratio is larger than 1). The current setup with a fastest and more sensitive camera as in~\cite{Cobelli} enables to have a higher precision of the measurement of $\Delta \omega$ and $C$ which must be the reason for the difference with the data shown in~\cite{Cobelli}. The curves at various $P$ are self similar and collapse to a master curve when plotted $\Delta \omega/P^{1/2}$ versus $k/P^{1/6}$ (fig.~\ref{shift}(b)). The scaling of $k/P^{1/6}$ corresponds through the LDR to the scaling $f/P^{1/3}$ that is also shown to collapse the time spectra~\cite{Boudaoud,Mordant}. It is due to the $P^{1/3}$ scaling of the cutoff frequency observed in the time spectra (fig.~\ref{specf}, \cite{Mordant}). The physical origin of the cutoff frequency is still unclear but it could be independent of dissipation as such a cutoff frequency can be exhibited by dimensional analysis of the Fˆppl-Von Karman equations (\ref{le})\&(\ref{nle})~\cite{Mordant}. The analytical expression of $\Delta \omega$ has not been derived yet from the WTT in the case of the elastic plate. Even if the WTT does not agree with the observed spectra, the scaling observed for $\Delta \omega$ may still be related to theoretical predictions. If one considers $\Delta \omega$ as a measure of the degree of non linearity, it is the strongest slightly above the forcing and decays to zero at high frequencies. Thus we do not expect the hypotheses of weak non linearity to breakdown at high frequencies as is the case in some systems~\cite{Newell}. The shift to the LDR is about $50\%$ of the LDR at the forcing scales and much smaller in the inertial range. Together with the absence of harmonics, this observation can be considered as a indirect validation of hypothesis (ii) of weak non linearity.

\subsection{The width of the dispersion relation}
\begin{figure}[!htb]
\vspace*{1cm}
\includegraphics[width=8.5cm]{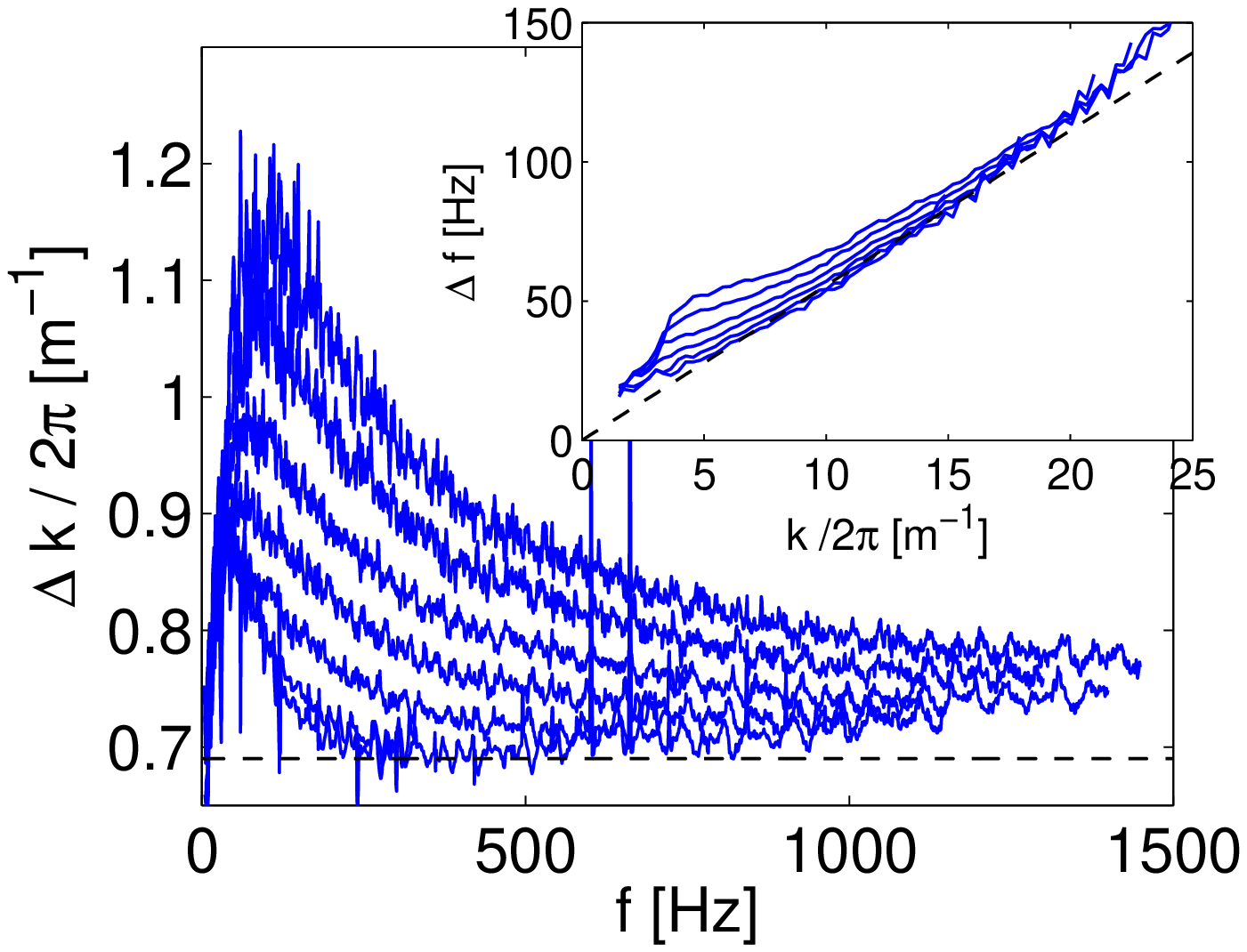}
\caption{The $k$-width of the energy line of fig.~\ref{sp2D} (see text) for $P=1$, 4, 9, 16, 25 and 36 (from bottom to top). The dashed line is the resolution of the space Fourier transform. The insert shows the $f$-width. The dashed line is the resolution of the space Fourier transform translated in frequency space (see text).}
\label{width}
\end{figure}
The theoretical correction $\Delta \omega$ can be complex {\it a priori}. Thus one expects that the width of the energy line around the NLDR in figure~{\ref{sp2D} should be increased by non linear effects due to the imaginary part of $\Delta \omega$~\cite{Newell}. This width has been computed in two ways in fig.~\ref{width}: either a $k$-width at constant $f$ or a $f$-width at constant $k$. The $k$-width has a lower bound due to the resolution of the space Fourier transform on the size of the picture. It is shown as a dashed line in fig.~\ref{width}. The measured $k$-width is indeed larger than this lower bound and is seen to increase monotonically with $P$. The lower bound of the $k$-width can be translated into a lower bound for the $f$-width through the group velocity of the LDR. It gives the dashed line in the insert of fig.~\ref{width} which is indeed a lower bound for the measured $f$-width. The width behaves similarly  to the shift to the LDR, with a peak slightly above the forcing wavenumber and a decay at the largest frequencies.

\section{Finite size effects}
\label{FSE}
\begin{figure}[!htb]
\includegraphics[width=9cm]{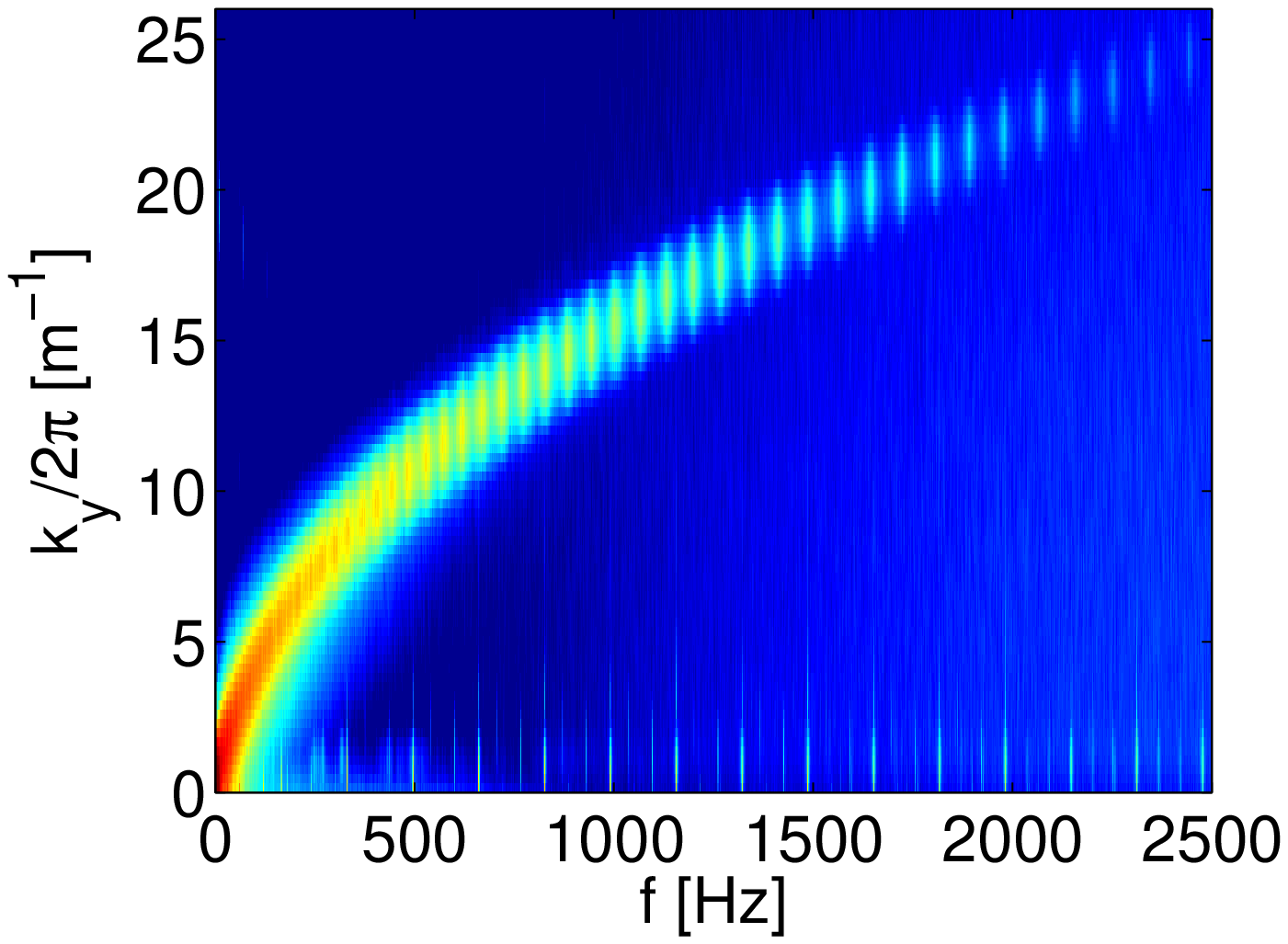}
\caption{Cut of the space time spectrum $E(\mathbf k,\omega)$ at $k_x=0$. Discrete frequencies are clearly visible.}
\label{fini}
\end{figure}
\begin{figure}[!htb]
\includegraphics[width=9cm]{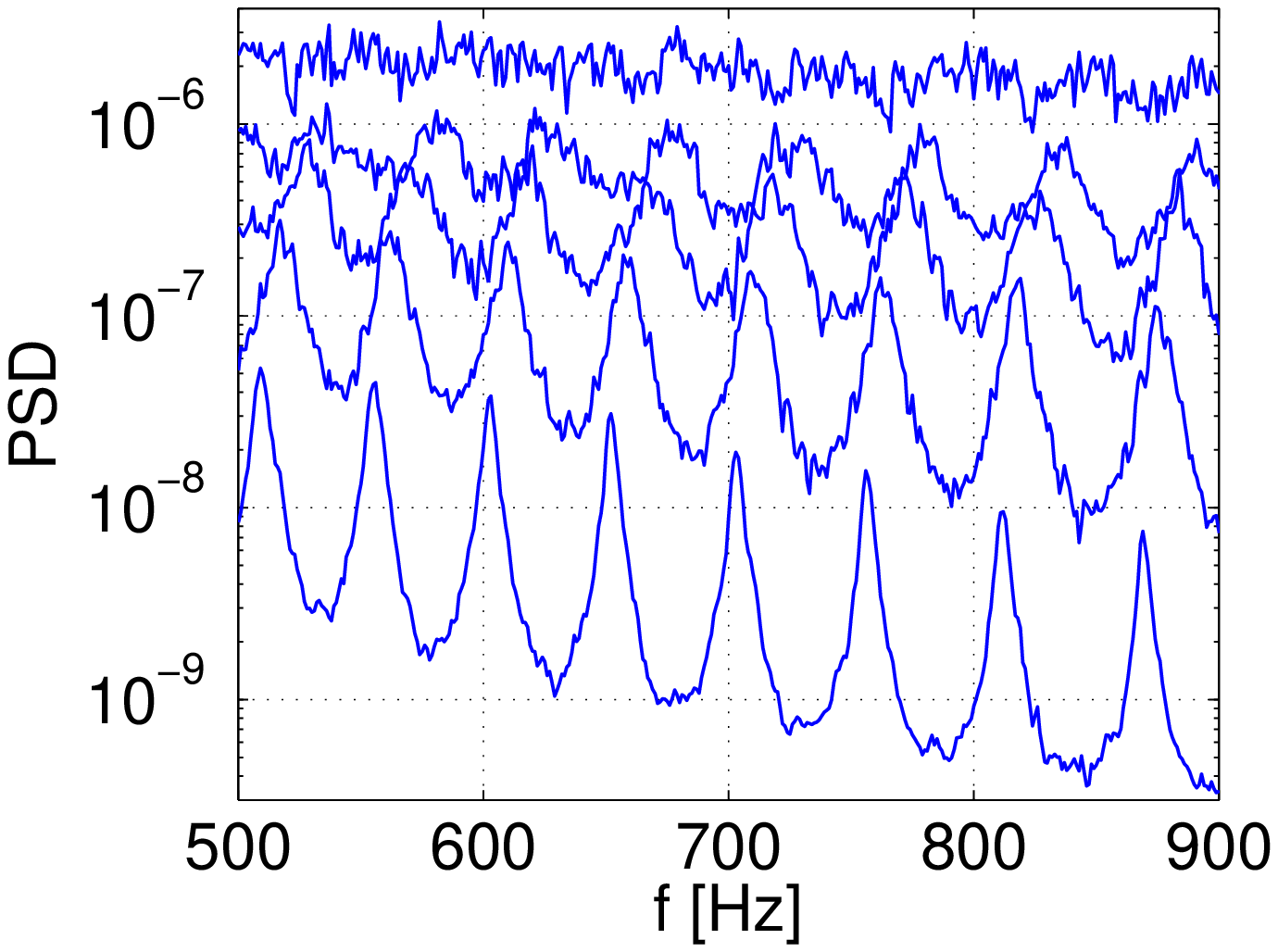}
\caption{Evolution of the spectral energy on the dispersion relation in the plane $k_x=0$ for $P=1$, 4, 9, 16 and 36 (from bottom to top). The curves have been shifted vertically by a factor $1.5$ for clarity.}
\label{fini1}
\end{figure}
The hypothesis (i) of WTT assumes an infinite system to avoid discrete values of the frequencies of the available vibration modes. Figure~\ref{fini} shows a cut of the spectrum $E(\mathbf k,\omega)$ in the plane $k_x=0$. At low frequencies, the energy line is continuous but at the highest frequencies the energy is localized in peaks. These peaks appear in the region of the frequency spectrum in which the spectrum is decaying exponentially and is the weakest. In this region the discreteness of the normal modes of the plate seem to appear. They are the most visible in the vicinity of the plane $k_x=0$ in the $(\mathbf k,\omega)$ space. The $y$ direction is along the short side of the plate. The separation of the modes is the widest along this direction. One can extract the amplitude of the energy line along the NLDR of fig.~\ref{fini}. The result is shown in fig.~\ref{fini1} for various forcing amplitudes and in the frequency window $f\in\left[500,900\right]$~Hz. The spectral width of the peak is increasing with the forcing until the peaks merge at the highest forcing. In the same frequency window, similar peaks can be seen also in other directions of $\mathbf k$ ($k_y=0$ for instance) but the interval between peaks is much lower so that the peaks merge for much smaller forcing amplitudes. This can be observed in the cuts of $E(\mathbf k,\omega)$ for given values of the frequency shown in fig.~\ref{spcut}. The two lower cuts show clear localized peaks of energy near $k_x=0$ (that are not due to insufficient convergence of the average). When $k_x$ increases the energy is continuously distributed along the circle that corresponds to the dispersion relation. 

Finite size effects are visible only in the high frequency, fast decaying region of the spectra. In this region, the measurement of the shift to the linear dispersion relation show that the non linearity is the weakest. The widening of the discrete frequencies is thus not strong enough to observe a continuous spectrum. In the power law region of the spectra, the energy is distributed continuously as the non linearity is stronger. 

In \cite{Nazarenko}, it was suggested that finite size effects could cause a ``sandpile behaviour" of the energy flux that eventually leads to a different scaling of the spectrum. The argument is the following: energy will be able to flow across scales only when the amplitude of the waves induces a widening of the frequencies of adjacent (in $k$-space) modes  equal to the separation of their frequencies. The frequency widening of the mode is of order $\frac{1}{n}\frac{dn}{dt}$ and it should be equal to the mode separation which is of order $\frac{\partial \omega}{\partial k}\delta k\sim k/L$ (with $L$ the size of the finite container which is of order 1~m here). From \cite{During} the collision integral scales as $\omega(k)^{1-3x}$ for a spectrum $n(k)\propto \omega(k)^{-x}$. The so-called ``critical spectrum" is then obtained by solving $\omega(k)^{1-2x}\propto \omega(k)^{1/2}$ which provides $x=1/4$. In \cite{During}, the collision integral is said to converge for $0.5<x<1.2$. Here we get $x=1/4$ which does not ensure the locality of the interactions in Fourier space (hyp. (iv)). The Kolmogorov-Zakharov solution corresponds to $x=1$ i.e. steeper than the possible critical spectrum. The observed spectrum is even steeper than the Kolmogorov-Zakharov prediction and thus not in agreement with the critical spectrum prediction. As the latter is affected by the non locality issue, this deserves more theoretical investigations. What is observed in fig.~\ref{fini} is that the non linear widening of the modes is greater than the mode separation at low frequencies so that a Kolmogorov-Zakharov cascade may proceed. When the widening is of the same order as the separation, avalanches may occur.  In the highest part of the spectrum, the modes are apparent and it must correspond to another phenomenology.

\section{Conclusions}
All measurements of wave turbulence in an elastic plate reported in this article are qualitatively consistent with the kinetic Weak Turbulence Theory. A transition from discrete wave turbulence to a kinetic wave turbulence is observed when the non linearity is sufficiently strong, as described by Kartashova \cite{Kartashova,Kartashova1,Kartashova2}, with a intermediate mesoscopic or laminated turbulence. The existence of a frequency cutoff leading to an exponential decay of the spectrum at large frequency may be due to this transition to discrete turbulence. It has been observed that the spectra are steeper in mesoscopic or discrete turbulence due to a limited set of available resonant frequencies.

The non linearity is weak in this system as quantified by the small shift of the observed non linear dispersion relation in respect to the LDR. As expected from the phenomenology, isotropy of the energy spectra is restored at large frequencies despite the anisotropic forcing. Despite this agreement with the hypotheses of the WTT, the energy spectrum is not in quantitative agreement with the theory. Recent work by Connaughton~\cite{Connaughton} have shown that in numerical simulation, the truncation scheme can induce a positive or a negative bottleneck effect that can be extended in a quite large interval of frequencies. The dissipative mechanism are complex and unknown to some extend in the case of the elastic plates. Nevertheless a negative bottleneck effect could be responsible for the quantitative disagreement observed here especially since the interval of accessible frequencies (let alone wavenumbers) is not extremely wide.

\thanks{The author thanks the french ``Agence Nationale de la Recherche" for its funding under grant TURBONDE BLAN07-3-197846.}
\bibliographystyle{epj}
\bibliography{plaqueiii}

 \end{document}